\documentclass[11pt]{article}

\usepackage[margin=1in]{geometry}
\usepackage{amsfonts,amssymb,amsmath,amsthm}
\usepackage{graphicx}
\usepackage{psfrag}
\usepackage{calc}
\usepackage{epic}
\usepackage[english]{babel}
\usepackage{hyperref}
\usepackage{float}
\usepackage{latexsym}
\usepackage{color}
\usepackage{subfigure,dsfont}

\allowdisplaybreaks

\usepackage{hyperref}
\newcommand{\footremember}[2]{%
    \footnote{#2}
    \newcounter{#1}
    \setcounter{#1}{\value{footnote}}%
}
 
\title{Stability conditions for a decentralised medium access algorithm: single- and multi-hop networks}
\author{%
  Seva Shneer\footremember{HW}{Heriot-Watt University, V.Shneer@hw.ac.uk}%
  \and Alexander Stolyar\footremember{UIUC}{University of Illinois at Urbana-Champaign, stolyar@illinois.edu}%
  }
\date{}
\newtheorem{theorem}{Theorem}
\newtheorem{lemma}[theorem]{Lemma}

\newtheorem{corollary}[theorem]{Corollary}

\theoremstyle{remark}
\newtheorem{remark}[theorem]{Remark}

\newcommand{\be}{ \begin{equation}}
\newcommand{\ee}{\end{equation}}
\newcommand{\ben}{ \begin{equation*}}
\newcommand{\een}{\end{equation*}}

\newcommand{\beql}[1]{\begin{equation}\label{#1}}
\newcommand{\eeql}{\end{equation}}
\newcommand{\eqn}[1]{(\ref{#1})}

\def\E{{\mathbb E}}
\def\P{{\mathbb P}}

\def\N{{\mathcal N}}
\def\C{{\mathcal C}}
\def\D{{\mathcal D}}

\begin{document}
\maketitle

\begin{abstract}
We consider a decentralised multi-access algorithm, motivated primarily by the control of transmissions in a wireless network. 
For a finite single-hop network with arbitrary interference constraints we prove stochastic stability under the natural conditions.
For infinite and finite single-hop networks, we obtain broad rate-stability conditions.
We also consider symmetric finite multi-hop networks and show that the natural condition is sufficient for stochastic stability.
\end{abstract}

\section{Introduction}

We consider a model motivated by wireless networks. A key feature of wireless transmissions is that they interfere with each other, especially if the receivers are in close proximity, and this interference may prevent some of the simultaneous transmissions from being received correctly. This creates the need for the design of algorithms regulating the behaviour of transmitters in wireless networks, so that simultaneous interfering transmissions do not occur at all, or occur rarely.

The transmitter-receiver pairs in a network are represented by nodes on a graph, and an edge between two nodes is present if the corresponding transmissions interfere with each other. Thus, the resulting interference graph represents the interference structure of the network.

We consider both single- and multi-hop networks. In a single-hop network jobs (or messages to be transmitted) arrive at nodes in a network and, upon a successful transmission, leave the network. The dependence between states of different nodes exists because the interference graph imposes constraints on simultaneous transmissions. In a multi-hop network, a message, upon successful transmission at one node, may leave the network or may move to another node, where it needs to be transmitted again. Thus, multi-hop networks add a further layer of complexity, as the states of the nodes are dependent not only due to interference constraints, but in this case also due to message movement between the nodes. 

We are interested in stability of a network. In finite networks by stability we mean the stochastic stability of the nodes' queues. One usually calls a transmission scheduling algorithm maximally stable if it guarantees stability if such is feasible at all, under at least one algorithm. The celebrated BackPressure (sometimes referred to as MaxWeight) algorithm introduced in \cite{Tassiulas1992} is maximally stable. It is however, in all but the simplest network stuctures, {\it centralised}, i.e. it requires the presence of a central entity that is aware of the state of the entire network. This is not practical in wireless networks that tend to be large and ever-changing.

There is thus the need for designing {\it decentralised} algorithms where each node regulates its behaviour on its own, without any global knowledge. In principle, decentralised behaviour may lead to conflicts, when several interfering transmissions will be attempted simultaneously, and the messages will not be received. 
We consider the so-called CSMA (Carrier Sense Multiple Access, see \cite{Kleinrock1975}) networks where each transmitter can sense if a neighbouring node is transmitting and will never initiate an interfering transmission. Conflicts are thus avoided in CSMA networks.

In this paper, it is not our goal to design decentralised algorithms that are maximally stable. We study a different question: what is the stability performance of some simple specific decentralised protocols? We consider the following protocol. Assume that the network is finite. Assume also that time is slotted, i.e. arrivals happen at discrete time instances denoted $1,2,\ldots$, all transmission times are equal to $1$, and transmissions start at the beginning of a unit-long time slot and complete at its end. Assume that at the beginning of each time slot, each message is assigned a random number, drawn independently from some fixed absolutely continuous distribution; the lower this number, the higher the message transmission priority in the slot. Then, a given message is transmitted in a slot, if and only if its priority is the highest among all messages within its neighbourhood, which includes its own node and all adjacent (neighbouring) nodes. 

In \cite{SnSt2017} a related protocol is considered, where nodes, rather than messages, compete for transmission, and where a node may transmit even if its neighbour has a higher priority, provided this neighbour does not transmit due to its other neighbour having an even higher priority. The protocol considered in this paper is significantly more conservative and thus, not surprisingly, leads to a smaller stability region.

On the face of it, the protocol described in the previous paragraph is centralised, as priorities need to be assigned. It is however easily implemented in a decentralised fashion (with an arbitrarily small loss of efficiency), for example as follows. Assume that each time slot is split into two parts: the first part, of duration $0 < \varepsilon < 1$, is devoted to medium-access competition, and the second part, of duration $1-\varepsilon$, is devoted to actual message transmissions. Assume that each message transmission lasts exactly $1-\varepsilon$. The protocol may be designed as follows: a random time uniformly distributed in $(0,\varepsilon)$ is chosen at the beginning of every time slot for every message currently present in the system, independently of all other messages, and these random variables are also independent over time slots. Once this time expires for a message, an access transmission is initiated. This access transmission does not start an actual transmission but is registered by all nodes in the neighbourhood. At time $\varepsilon$ all access transmissions stop. If a node registered that one of its messages' access transmission was the earliest in its neighbourhood, the node will transmit that message. It is easy to see that the loss of efficiency versus the centralised protocol with transmission times of $1$ is exactly $\varepsilon$.

In the rest of the paper, for simplicity we consider a ``cleaner'' version of the protocol, as described in the previous paragraph that ignores the loss of an $\varepsilon$ proportion of the throughput.

The following example illustrates the conservative nature of the protocol.
Consider $4$ nodes, with interference graph being a ``circle'', so that either nodes $1$ and $3$, or nodes $2$ and $4$ can transmit simultaneously. Assume that, in a given time slot, the message with the highest priority is located at node $1$, the message with the second-highest priority is located at node $2$ and the message with the third-highest priority is located at node $3$. Under the algorithm considered here, only the first message will be transmitted in this slot, when in fact nodes $1$ and $3$ could successfully transmit simultaneously. Of course, under other priority orderings, transmission of two messages will occur.

Denote by $\mathcal{N}_i$ the neighbourhood of node $i$ in the interference graph (all neighbours of the node and the node itself). Denote by $X_i$ the number of messages at node $i$ at the beginning of a time slot. It is easy to see that node $i$ will transmit a message with probability
\begin{equation} \label{eq:rate}
\varphi_i= \frac{X_i}{\sum_{j \in \N_i} X_j}.
\end{equation}

A model closely related to ours has been considered in a recent paper \cite{Baccelli2018} (see also \cite{Baccelli2017}). It is a single-hop network with nodes located on a grid. The model is in {\em continuous time}, with each message having an exponentially distributed size with unit mean. {\em All} messages may transmit simultaneously, with the instantaneous transmission rate depending on the interference from the messages in the neighbourhood. A standard assumption that a message transmission rate is proportional to its Signal-to-Noise-Ratio is adopted in \cite{Baccelli2018}, which leads to a {\em node} $i$ transmission rate given exactly by \eqn{eq:rate}. The model in \cite{Baccelli2018} is {\em symmetric} in that the message arrival rates at all nodes are equal. The authors focus on {\it infinite-grid} networks and are interested in their stability. The authors define this as the finiteness of the minimal stationary regime for the system starting with all queues being empty (see \cite{Baccelli2018} for more details). They show that a network is stable in this sense under the natural condition on the message arrival rates. The main tool in their analysis is {\it monotonicity}, i.e. the property that if one network starts with an initial condition dominating that of another, there exists a coupling preserving this dominance at all times.

In the single-hop scenario, we consider {\em arbitrary} networks: finite or infinite, arbitrary interference graph, arbitrary arrival intensities. For finite single-hop networks we prove that the system is stochastically stable if the arrival rates belong to a certain set. For infinite single-hop networks we obtain a rather broad sufficient condition for the {\em rate-stablity} which is the property that, starting from any fixed initial state, the growth rates of the queues are sub-linear in time.

We also consider finite {\em multi-hop} networks; here we additionally assume that a network is symmetric: it is a regular graph (recall that a graph is regular if all nodes have the same number of neighbours), with equal exogenous arrival rates at the nodes and with a message path through the network being the standard random walk (until the message leaves the network). In the multi-hop setting, a further complication arises from the fact that monotonicity does not hold. We prove directly that the network is stochastically stable under a natural condition ensuring that, when states of all queues on the network are equal, the average rate at which work arrives is smaller than the average rate at which work is performed by the system. Our approach is {\em not} based on monotonicity, in either setting.

Our stability proofs use the fluid-limits technique. The discrete-time setting motivating our work and the continuous-time network motivating \cite{Baccelli2018} share the same fluid limits, and thus our results are valid in the continuous-time setting too. Note also that the random variables representing the number of successful transmissions from all nodes in the same time slot in our model are not independent. However, our stability results also apply to a different discrete-time model, where 
 (perhaps rather unrealistically) each station transmits a message with a probability given by \eqref{eq:rate}; again, this is due to the fact that this model has the same fluid-limit dynamics as ours.

Another important concept in wireless networks is utility maximisation. Utility-optimal algorithms are known to guarantee maximal stability for finite single-hop networks, under some assumptions on the utility functions. However, these algorithms are centralised as the {\em average} service rates $\varphi_i$ assigned to nodes form a solution $\overline{\varphi} = \{\varphi_i\}$ to a global optimisation problem.  An important example of such algorithms is presented by the well-known $\alpha$-fair algorithms (see \cite{Kelly1998,Mo2000,Roberts2000} for introduction of the fair-allocation concepts and \cite{Bonald2001,DeVeciana2001} for stability proofs).  In $\alpha$-fair algorithms the average rates $\varphi_i$ are such that
$$
\overline{\varphi} \in \arg\max_{\overline{\mu} \in \C} \sum_i X_i \frac{1}{1-\alpha} \left(\frac{\mu_i}{X_i}\right)^{1-\alpha}, ~~\mbox{when}~~\alpha > 0, ~\alpha \ne 1,
$$
or
$$
\overline{\varphi} \in \arg\max_{\overline{\mu} \in \C} \sum_i X_i \log (\mu_i/X_i), ~~\mbox{when}~~\alpha = 1,
$$
where $\C$ is some fixed set. 
In the finite single-hop setting, for a specific (natural) set $\C$,
 we provide two proofs of stochastic stability, for the arrival rates within  $\C$.
The first proof follows from a much more general result for monotone $0$-homogeneous service rates, which is of interest on its own and may have other applications. Our infinite-system rate stability proof is based on the same ideas. We provide a second proof for finite systems, as it 
is based on discovering an important property of the rates \eqref{eq:rate}: they in fact happen to be $\alpha$-fair in $\C$, with $\alpha=2$. We believe this property to be interesting in its own right too as it presents an example of a decentralised protocol which happens to be centrally optimal in a certain sense, and it is known to imply stochastic stability in finite single-hop networks.

It is known that utility-maximising algorithms, with the exception of proportionally - fair algorithms (i.e. $\alpha$-fair algorithms with $\alpha=1$; see \cite{Walton2015,Bramson2017} for a treatment of some of the multi-hop scenarios), in general do not guarantee stability in multi-hop networks. Therefore our result on the fairness of the rates \eqref{eq:rate} does not imply stability in the multi-hop setting. In Section \ref{sec:multi-res} we 
present our main result in the finite multi-hop setting stating that a natural stability property does hold under our algorithm for a class of symmetric networks. Specifically, these networks are such that the interference graph is regular, the exogenous arrival rates to all nodes are equal, and upon a successful transmission a message may either leave the network or move to a neighbour node chosen uniformly at random. We show stochastic stability under a natural condition on the per-node exogenous arrival rate.


The paper is organised as follows. We define our model in Section \ref{sec:model} and then present our main results in Section \ref{sec:results} (for single-hop networks in Section \ref{sec:single-res} and for multi-hop networks in Section \ref{sec:multi-res}). Section \ref{sec:fluid} contains the necessary fluid limit constructions, definitions and results.
The proofs of our main results are presented in Section \ref{sec:single} for the finite single-hop case, in Section \ref{sec-infinite} for the infinite single-hop case, and in Section \ref{sec:multi} for the multi-hop case. We discuss some open problems in Section \ref{sec:open}.

{\bf Basic notation.} We will use the following notation throughout: $\mathbb{R}$ and $\mathbb{R}_+$ are the sets of real and real non-negative numbers, respectively;
$\overline{y}$ means (finite- or infinite-dimensional) vector $(y_i)$; for a finite-dimensional vector $\overline{y}$, $\|y\| = \sum_i |y_i|$;
for a set of functions $(f_i)$ and a vector $(y_i)$, $\overline{f}(\overline{y})$ denotes the vector $(f_i(\overline{y}))$; vector inequalities are understood component-wise;
$\frac{d^+}{dt}$ is the right derivative; $\frac{d_l^+}{dt} y(t_0) = \lim \inf_{t \downarrow t_0} \frac{y(t)-y(t_0)}{t-t_0}$ -- the lower right Dini derivative; $y(\cdot) = (y(t),~t\ge 0)$; we also use the convention that $0/0 = 0$. The indicator function of an event or condition $A$ is denoted by $I(A)$; $\lfloor a \rfloor$ is the largest integer not exceeding $a$.
Abbreviation {\em r.v.} means {\em random variable}; {\em w.p.1} means {\em with probability $1$}; {\em i.i.d} means {\em independent identically distributed}; {\em u.o.c.} means {\em uniformly on compact sets}. 

\section{Model and notation} \label{sec:model}

Denote by $\mathcal{V}$ the set of nodes of a graph, and by $\mathcal{E}$ the set of its edges. The set $\mathcal{V}$ may be finite (in which case we will refer to the network as finite) or countable (in which case we will refer to the network as infinite). Denote by $N$ the (finite or infinite) cardinality of $\mathcal{V}$.

For a node $i$, denote by $\N_i = \{j \in \mathcal{V}: (i,j) \in \mathcal{E}\} \cup \{i\}$ its neighbourhood. We assume throughout that $\N_i$ is finite for each $i$, the graph $\mathcal{G} = (\mathcal{V}, \mathcal{E})$ is connected, and that the neighbourhood relationship is symmetric (or that the graph is undirected), i.e. if $i \in \N_j$, then $j \in \N_i$.

Each node has an infinite buffer for storing messages but there is no queue. Time is slotted, and at the beginning of each time slot first transmissions are initiated, and then arrivals happen. Each transmission time is equal to $1$.

At the beginning of each time slot, every message in the system is assigned a random number which is drawn, independently of everything else, from a certain fixed absolutely continuous distribution; the smaller this number, the higher the message transmission priority.
A message is transmitted if and only if it has the highest priority in its neighbourhood, i.e. node $i$ transmits a message if that message's priority is the maximal over all the messages in $\N_i$. We refer the reader to the introduction for an explanation of how this may be implemented in a decentralised way, by using Uniform distribution on a small interval, with an arbitrarily small loss of throughput.

At each time slot $k$, a random number $\xi_i(k)$ new messages arrive at node $i$. We assume that $\xi_i(k)$ are i.i.d. with $\E(\xi_i(k)) = \lambda_i>0$.

Throughout the paper we use notation 
\begin{equation} 
\label{eq:varphi}
\varphi_i(\overline{p}) = \frac{p_i}{\sum_{j \in \N_i} p_j}, ~~ \overline \varphi(\overline{p}) = (\varphi_i(\overline{p})),
\end{equation}
where $\overline{p} = (p_i)$ is a vector with finite non-negative components. By convention, $\varphi_i(\overline{p})=0$ when 
$\sum_{j \in \N_i} p_j =0$.

\section{Main results} \label{sec:results}

The results are split into two subsections covering single- and multi-hop networks. The subsection on single-hop networks contains results for both finite and infinite networks.

\subsection{Single-hop network} \label{sec:single-res}

For the single-hop network we consider any undirected graph $\mathcal{G}$ where each node has a finite neighbourhood. We assume that, upon a successful transmission, a message leaves the system. The evolution of the state of the queue of node $i$ may then be written as
$$
X_i(k+1) = X_i(k) + \xi_i(k) - \eta_i(k),
$$
where by $X_i(k)$ we denoted the state of the queue of node $i$ at time $k$, and by $\eta_i(k)$ - the number of messages leaving node $i$ during the $k$-th time slot. The random variable $\eta_i(k)$ can only take values $0$ and $1$, and it is easy to see that, as priorities are chosen independently from the same fixed distribution,
$$
\P(\eta_i(k)=1|\overline{X}(k) = \overline{X}) = \varphi_i(\overline{X}) = \frac{X_i}{\sum_{j \in \N_i} X_j}.
$$
Denote
\begin{equation} \label{eq:stab_set_csma}
\C = \left\{\overline{\lambda}: \overline{\lambda} \le \overline{\varphi}(\overline{p}) \quad \text{for some} \quad \overline{p} \in \mathbb{R}_+^{N} \right\}.
\end{equation}

We will call a finite network {\em stochastically stable} if the countable Markov chain $\overline X(\cdot)$ is positive recurrent.

\begin{theorem} \label{thm:stab_csma_single}
Consider a finite network. If $\overline{\lambda} < \overline{\nu}$
for some $\overline{\nu} \in \C$, then the system is stochastically stable.
\end{theorem}

The fact that $\bar{\lambda}$ is component-wise strictly smaller than a vector from $\C$ means that there exists at least one configuration of queue states, represented by the corresponding vector $\bar{p}$ from \eqref{eq:stab_set_csma}, such that for that configuration the amount of work arriving at any node is strictly smaller than the average amount of work this node performs. This is a natural sufficient stability condition. We conjecture that this condition is essentially necessary as well,
in that if $\overline{\lambda}$ is strictly outside $\C$, the process is transient. However, establishing this fact is beyond the scope of the present paper -- it may be a subject of future work.


We present a proof of Theorem \ref{thm:stab_csma_single} in Section \ref{sec:single}.

\begin{corollary} \label{cor:1}
In a finite symmetric network, where the graph $\mathcal{G}$ is $(m-1)$-regular (so that each node has degree $(m-1)$) and $\lambda_i = \lambda$ for each $i$, the condition of Theorem \ref{thm:stab_csma_single} is equivalent to the requirement that $\lambda < 1/m$. In particular, for a network of nodes located on a circle with the same arrival intensity $\lambda$ at each node, $\lambda < 1/3$ guarantees stability.
\end{corollary}

The example of a network of nodes located on a circle illustrates well the fact that, due to the conservative nature of the algorithm, its stability set, $\lambda < 1/3$, is smaller than the maximal stability set $\lambda < 1/2$ achieved by, for instance, MaxWeight. It is also smaller than the set $\lambda < 2/5$ achieved by the less conservative protocol considered in \cite{SnSt2017}.

We now prove corollary \ref{cor:1} Indeed, if $\lambda < 1/m$, then the vector $(\lambda,\ldots,\lambda)$ is component-wise upper-bounded by the vector $(1/m,\ldots, 1/m)$ which belongs to the set $\C$ trivially (one needs to take the vector $\overline{p} = (1,\ldots,1)$ to verify this). Assume now that for some $\bar p$ the vector $(\lambda,\ldots,\lambda)$ is component-wise smaller than a vector $\bar \varphi (\bar p) \in \C$. Then 
$$
\frac{1}{\lambda} > \sum_{j \in \N_i} \frac{p_j}{p_i}
$$
for each $i$, and if we add up these inequalities, we obtain
$$
\frac{N}{\lambda}  > \sum_i \sum_{j \in \N_i} \frac{p_j}{p_i} = \frac{1}{2} \sum_i \sum_{j \in \N_i} \left(\frac{p_j}{p_i} + \frac{p_i}{p_j}\right) \ge mN,
$$
which implies $\lambda < 1/m$.

\begin{remark} \label{remark:continuous}
As our proof of Theorem \ref{thm:stab_csma_single} is based on fluid limits, its results are also valid for a continuous version of the model similar to that of \cite{Baccelli2018}. We refer the reader to the Introduction for an explanation of the connection between models.
\end{remark}

A finite or infinite network is called {\em rate-stable}, if w.p.1, 
$$
\lim_{k\to\infty} X_i(k)/k = 0, ~~~\forall i,
$$
for any initial state $\overline{X}(0)$ with all components being finite, $X_i(0)<\infty$.

Rate-stability is a weaker property than the stochastic stability. The following result gives a sufficient condition for rate-stability.

\begin{theorem} 
\label{thm:rate-stab}
The infinite or finite system is rate-stable if $\bar \lambda \le \bar\varphi(\bar p)$ for some $\bar p$ such that $0 < c \le p_i \le  1$ for all $i$.
\end{theorem}

A proof of Theorem~\ref{thm:rate-stab} is given in Section~\ref{sec-infinite}. 

\subsection{Symmetric multi-hop networks with Geometric service requirements} \label{sec:multi-res}

In this section we restrict our attention to a finite $(m-1)$-regular graph, $m \ge 2$. Assume that the access procedure is the same as before. Now however, upon service, a message leaves the system with probability $1/k$, and goes to a neighbouring node with probability $(1-1/k)\frac{1}{m-1}$. Assume that arrival rate into each node is $\lambda/k$, so that the total workload for each node is $\lambda$ -- this follows from standard rate-balance equations.

One can think of each message needing a Geometric$(1/k)$ number of successful transmissions to leave the system and, conditionally on not leaving the system upon a successful transmission, performing a simple random walk on the graph, i.e., choosing a neighbouring node uniformly at random.

\begin{theorem} 
\label{thm:stability_csma_multi}
Suppose the system graph is finite $(m-1)$-regular, $m \ge 2$. Then,
if $\lambda < 1/m$, the system is stochastically stable.
\end{theorem}

We provide a proof of Theorem \ref{thm:stability_csma_multi} in Section \ref{sec:multi}.

\begin{remark}
Similarly to Remark \ref{remark:continuous}, our results in the multi-hop setting hold for a continuous-time version of the model.
\end{remark}

\section{Fluid limits} \label{sec:fluid}

\subsection{Fluid scaling}
\label{sec-fluid-scaling}

Our main results are based on the fluid-limit technique (see \cite{RybSt92,Dai95,St95}). For the application of this technique to {\em discrete time} processes, see e.g. \cite{CS2001,St2005alq}. 

We consider the Markov chain $\overline{X}(k), ~k=0,1,2,\ldots,$ and extend it to continuous time with the convention $\overline{X}(t) = \overline{X}(\lfloor t \rfloor)$. Consider a sequence of processes $\overline{X}^{(r)}(\cdot)$, indexed by $r \uparrow \infty$, and their {\em fluid-scaled} versions
$$
\overline{x}^{(r)}(t) = \frac{\overline{X}^{(r)}(rt)}{r}, ~~t\ge 0.
$$

\subsection{Fluid sample paths}
\label{sec-fsp}

Fluid sample paths (FSP) are defined as possible limits of the realisations of $\overline x(\cdot)$, with common ``driving'' processes' realisations, satisfying the functional strong law of large numbers. The definition is along the same lines as FSP definitions in other contexts (see e.g. the proof of \cite[Theorem 2]{St2005alq} for a context somewhat similar to ours). However, it is not quite standard, for two reasons. First, the service process construction (i.e., the procedure determining the transmission schedule in each slot) is not quite standard. Second, we need to define FSPs with possibly countable set of component functions. For these reasons we give a formal definition here. To improve the exposition, the definition will be for the single-hop case only. Extension to the multi-hop case is straightforward (we will comment on that later).

We start with specifying a construction of the service process,
 which is consistent with the model definition. 
Recall that the neighborhood $\N_i$ of each node $i$ is finite. Suppose the nodes are indexed by $i=1,2,\ldots$. (This indexing is arbitrary.)
Given the queue lengths $\overline X$ in a given time slot $k$, the transmission schedule in this slot is determined recursively as follows. Suppose, the mutual ranking $m(j) \in \mathcal{M}_j$ of the first $j$ nodes is already determined; here $\mathcal{M}_j$ is the set of all $j!$ permutations of $(1,\ldots,j)$.  The smaller the number (ranking level) of a node, the higher the ranking.
(Note that for $j=1$, the ``mutual'' ranking $m(1)$ of the single node $1$ is deterministic, 
trivially equal to $(1)$.)
The following describes how the mutual ranking $m(j+1) \in \mathcal{M}_{j+1}$ of the first $j+1$ nodes is determined. 
Given $m(j)$, the ranking $m(j+1)$ is random, but it is consistent with $m_j$. Namely, node $j+1$ gets a random ranking level $n$ in $\{1,\ldots,j+1\}$, and those nodes in $\{1,\ldots,j\}$, whose ranking level in $m(j)$ was greater or equal to $n$, will have their ranking level 
shifted up by 1. The random ranking level $n$ of node $j+1$ is determined as follows.
Associated with each index $j$ and each ranking (permutation) $m\in \mathcal{M}_j$ and each time $k$, there is an independent  r.v. $\kappa(j,m,k)$, uniformly distributed in $[0,1)$. Let $\pi_n$, $n=1,\ldots, j+1$, be the conditional on $(X_1, \ldots, X_j, X_{j+1})$ and the mutual ranking $m\in \mathcal{M}_j$ of the nodes $1,\ldots,j$,
probabilities that node $j+1$ receives the mutual ranking level $n$ among nodes $1,\ldots,j+1$. Then, node $j+1$ receives mutual ranking level $n$ if
$$
\kappa(j,m,k) \in [\sum_{\ell=1}^{n-1} \pi_\ell, \sum_{\ell=1}^{n} \pi_\ell) 
$$
Note that for any node $i$ there exists a finite $j$ such that $\N_i \in \{1,\ldots,j\}$. Therefore, even though this mutual ranking determination procedure has infinite number of steps, i.e. it is a (random) sequence mutual rankings $m(1), m(2), \ldots$, 
the mutual rankings within $\N_i$ -- and therefore the decision on whether or not $i$ transmits -- will be determined within finite number of steps.
We see that the service process is driven by a countable set of i.i.d. r.v. $\kappa(j,k,m)$ for all $k$, $j$, and $m\in \mathcal{M}_j$.

The driving processes for the arrivals are natural. Namely, a countable set of r.v. $\xi_i(k)$ for all $k$ and $i$, giving the number of arrivals in node $i$ at time $k$.
The r.v. $\xi_i(k)$ are independent across all $k$, and identically distributed across $k$ for each $i$.

The driving processes are easily seen to satisfy the following Functional Strong Law of Large Numbers (FSLLN) properties. W.p.1., 

$\forall j$ and $\forall m\in \mathcal{M}_j$, 
\beql{eq-lln1}
\lim_{r\to\infty} (1/r) \sum_{k \le rt} I\{\kappa(j,k,m) \le u\} = ut, ~~\mbox{u.o.c. in $(u,t) \in [0,1] \times [0,\infty)$}, 
\eeql

and, $\forall i$,
\beql{eq-lln2}
\lim_{r\to\infty} (1/r) \sum_{k \le rt} \xi_i(k)= \lambda_i t, ~~\mbox{u.o.c. in $t \in [0,\infty)$}.
\eeql

We can and do assume that
all processes $\bar X^{(r)}(\cdot)$ (for all $r$), and their fluid-scaled versions $\bar x^{(r)}(\cdot)$, are constructed on a common probability space,
defined by the driving processes $\{\kappa(j,k,m)\}$ and $\{\xi_i(k)\}$.

A vector-function
$\overline{x}(t) = (x_i(t)), ~t\ge 0$ is called a {\em fluid sample path} (FSP), with initial state $\overline{x}(0)$ having finite components $x_i(0)<\infty$,
if there exists a sequence $r\to\infty$ and a {\em realization} of the driving processes $\{\kappa(j,k,m)\}$ and $\{\xi_i(k)\}$, such that the FSLLN
conditions \eqn{eq-lln1} and \eqn{eq-lln2} hold and the corresponding sequence of fluid-scaled process {\em realizations} is such that
\beql{eq-fsp-def}
x_i^{(r)}(t) \to x_i(t), ~~\mbox{u.o.c. in $t \in [0,\infty)$, ~$\forall i$}.
\eeql

The above FSP definition is for the single-hop network, where each transmitted message leaves the system. Extending this definition to a multi-hop case is straightforward. An additional driving process $\zeta_i(\ell)$ consisting of uniformly distributed in $[0,1)$ r.v. is defined: $\zeta_i(\ell)$ determines the routing of the $\ell$-th message departing node $i$. This driving process satisfies the FSLLN, analogous to \eqn{eq-lln1}. An FSP is defined, again, as a limit of fluid-scaled trajectories corresponding to the driving processes' realizations satisfying the FSLLN conditions. We omit further details.

We emphasize again that the set of FSP components $x_i(\cdot)$ is finite for a finite system and countable for an infinite one. In either case, it is easy to see that all FSP components $x_i(\cdot)$ are Lipschitz {\em uniformly in $i$}.

\subsection{Fluid limit result}
\label{sec-fl-result}

\begin{lemma}
\label{lem-fl}
For either the single- or the multi-hop setting, the following holds. 
Suppose for a given sequence $r\to\infty$ the initial states of the fluid-scaled process are such that $\limsup_{r\to\infty} x_i^{(r)}(0) < \infty$ for each $i$.
Then, w.p.1., any subsequence of $\overline{x}^{(r)}(\cdot)$ contains a further subsequence such that 
$x_i^{(r)}(t) \to x_i(t)$, u.o.c., for each $i$, where $\overline x(\cdot)$ is an FSP.
\end{lemma}

The proof of Lemma~\ref{lem-fl} is very standard and is omitted here. For example, it can follow the exact same lines as that of \cite[Theorem 2]{St2005alq}, for a model close to ours.


\subsection{Stability}

For a {\em finite} system, to establish stochastic stability (positive recurrence) of the Markov chain $\{\overline{X}(k)\}_{k\ge 0}$, it suffices to prove (see \cite{RybSt92}) that for some $T>0$ and $\varepsilon>0$ any sequence of processes  $\overline{X}^{(r)}(\cdot)$, with $\|\overline{X}^{(r)}(0)\|=r$, is such that 
$$
\limsup_{r\to\infty} \E \frac{1}{r} \|\overline{X}^{(r)}(rT)\| \le 1-\varepsilon.
$$
It is a standard result when applying fluid-limit technique \cite{RybSt92,Dai95,St95}, that for the above to hold, it is sufficient to show that for some $\varepsilon>0$ and $T>0$, any FSP with $\|x(0)\|=1$ is such that
\begin{equation} 
\label{eq:fluid_stability}
||\overline{x}(T)|| \le 1-\varepsilon.
\end{equation}

For a {\em finite or infinite} system, to show rate-stability, it suffices to prove 
(see Lemma~\ref{lem-infinite} below)
that any FSP starting from zero initial state (all $x_i(0)=0$), stays in zero state at all times, $x_i(t)=0, ~t\ge 0$.
This follows from the rate-stability definition and Lemma~\ref{lem-fl}.

\section{Single-hop network} \label{sec:single1}

This Section is devoted to the proofs of our main results for single-hop networks. In Section \ref{sec:fsp} we establish further properties of fluid sample paths. Section \ref{sec:single} is devoted to the proof of Theorem \ref{thm:stab_csma_single} and Section \ref{sec-infinite} is devoted to the proof of Theorem \ref{thm:rate-stab}.

\subsection{FSP properties} \label{sec:fsp}

We start by establishing properties of the FSP dynamics in Lemma \ref{lemma:fluid_single}.

\begin{lemma} \label{lemma:fluid_single}
Any FSP 
in the single-hop case satisfies the following conditions:
\beql{eq-sh1}
x_i(t) > 0 \quad \text{implies} \quad  x'_i(t) = \lambda_i - \varphi_i(\overline{x}(t)), ~~~\mbox{for almost all $t\ge 0$},
\end{equation}
\beql{eq-sh2}
[x_i(t) = 0 ~\mbox{and} ~\sum_{j \in \N_i} x_j(t) > 0]  \quad \text{implies} \quad \frac{d^+}{dt} x_i(t) = \lambda_i.
\end{equation}
In particular, by property \eqn{eq-sh2}, any FSP is such that, for a fixed $i$, $\sum_{j\in \mathcal{N}} x_i(t) > 0$  implies that
$x_i(\tau) > 0$ for all $\tau>t$ sufficiently close to $t$. Furthermore, if the network is finite,
any FSP is such that $\sum_i x_i(t) > 0$ implies that
$x_i(\tau) > 0, ~\forall i,$ for all $\tau>t$ sufficiently close to $t$.
\end{lemma}

A proof of Lemma~\ref{lemma:fluid_single} may be given following the exact same lines as that of \cite[Theorem 2]{St2005alq}, for a model close to ours, and we omit it here.

%

\subsection{Proof of Theorem \ref{thm:stab_csma_single}}
\label{sec:single}

We present two different proofs of Theorem \ref{thm:stab_csma_single}. 
One proof of \eqn{eq:fluid_stability}, and then of Theorem~\ref{thm:stab_csma_single}, follows from the following much more general result.

For a function $\overline{\psi}=\overline{\psi}(\overline p)$, mapping a finite-dimensional positive orthant $\mathbb{R}^N_+$, $N < \infty$, 
into itself,
define
\begin{equation} \label{eq:stab_set_general}
\D = \left\{\overline{\lambda} \in \mathbb{R}^N_+: \overline \lambda \le \overline \psi(\overline{p}) \quad \text{for some} \quad \overline{p} \right\}.
\end{equation}

\begin{lemma} \label{lemma:stab_general}
Consider a family of Lipschitz trajectories $\overline x(t), ~t\ge 0$, in $\mathbb{R}^N_+$, $N < \infty$,
which satisfy the following conditions:
\beql{eq:FSP1}
x_i(t) > 0 \quad \text{implies} \quad  x'_i(t) = \lambda_i - \psi_i(\overline{x}(t)), ~~\mbox{for almost all $t\ge 0$},~~\mbox{for any $i$},
\end{equation}
\beql{eq:FSP2}
\sum_i x_i(t) > 0  \quad \text{implies} \quad \mbox{$x_i(\tau) > 0,$ for all $i,$ for all $\tau>t$ sufficiently close to $t$},
\end{equation}
where the function $\overline{\psi}$ is such that:

{\it (A)} each $\psi_i$ is non-increasing in $x_j$ for all $j \neq i$;

{\it (B)} each $\psi_i$ is $0$-homogeneous, i.e. $\psi_i(s \overline{x}) = \psi_i(\overline{x})$ for all $s > 0$ and for all $\overline{x}$.

Assume that $\overline{\lambda}$ is such that $\overline{\lambda} < \overline{\nu}$ for some $\overline{\nu} \in \D$. 
Then for any constants $0<\delta < K < \infty$, there exists $T>0$ such that, for any such trajectory with $\|x(0)\| = K$,
$$
\|x(T)\| \le \delta.
$$
\end{lemma}

{\bf Proof of Lemma \ref{lemma:stab_general}.} We now prove Lemma \ref{lemma:stab_general}. Fix a vector $\overline{p}$ such that 
$$
\overline \nu \le \overline \psi(\overline{p})
$$
for every $i$. Note that as $\lambda_i > 0$ for each $i$, necessarily $p_i > 0$ for each $i$.  Consider the function
$$
F(\overline{y}) = \max_i \left(\frac{y_i}{p_i}\right).
$$
For ease of notation, in the rest of the proof we drop the index $t$ and make the dependence of $\overline{\psi}$ on $\overline{x}(t)$ implicit.

Denote
$$
\mathcal{K} = \left\{k: k \in \arg \max_i \left(\frac{x_i}{p_i}\right)\right\}.
$$
The function $\max_i \{x_i/p_i\}$ is Lipschitz, because all $x_i(\cdot)$ are Lipschitz. The time points $t$, where the derivatives of all $x_i$ and of
 $\max_i \{x_i/p_i\}$ exist, are called regular. Almost all points (with respect to Lebesgue measure) are regular. Then, 
 due to \cite[Lemma 2.8.6]{Dai1999}, $\left(\frac{x_k}{p_k}\right)' = \left(\frac{x_l}{p_l}\right)'$ at any regular point of $F$ for any $k, l \in \mathcal{K}$. The derivative of the function is $F$ at a regular point is then
$$
(F(\overline{x}))' = \frac{1}{p_k} (\lambda_{k} - \psi_k)
$$
with an arbitrary $k \in \mathcal{K}$. Note that, as $\frac{x_k}{p_k} \ge \frac{x_j}{p_j}$ for any $k \in \mathcal{K}$ and for any $j$, due to property {\it (A)},
\begin{align*}
\psi_k & = \psi_k(\overline{x}) \ge \psi_k\left(\frac{x_k p_1}{p_k}, \ldots, \frac{x_k p_{k-1}}{p_k}, x_k, \frac{x_k p_{k+1}}{p_k}, \ldots, \frac{x_k p_N}{p_k}\right)
 \\ & = \psi_k\left(\frac{x_k p_1}{p_k}, \ldots, \frac{x_k p_{k-1}}{p_k}, \frac{x_k p_k}{p_k}, \frac{x_k p_{k+1}}{p_k}, \ldots, \frac{x_k p_N}{p_k}\right)
\\ & = \psi_k(\overline{p}) \ge \nu_k
\end{align*}
where in the last step we used property {\it (B)}.

Noting that there exists $\varepsilon > 0$ such that $\lambda_i < \nu_i - \varepsilon$ for every $i$, we obtain
$$
(F(\overline{x}))' = \frac{1}{p_k} (\lambda_{k} - \nu_k) + \frac{1}{p_k} (\nu_{k} - \psi_k) < -\frac{\varepsilon}{p_k}.
$$
This implies that $(F(\overline{x}))' $ is negative, bounded away from $0$, as long as $F(\overline{x})$ is positive, bounded away from $0$. This  concludes the proof of Lemma \ref{lemma:stab_general}. \qed

Theorem \ref{thm:stab_csma_single} follows from Lemma \ref{lemma:stab_general} as the rates $\overline{\varphi}$ clearly satisfy conditions {\it (A)} and {\it (B)}.

\begin{remark}
Lemma \ref{lemma:stab_general} is rather general and relates to the so-called cooperative dynamical systems (see, e.g. \cite{Hirsch1985,Smith1995}). We believe that this result is interesting on its own and may have other applications. This result also allows, in an obvious fashion, to obtain stability conditions for networks with a more general notion of neighbourhood considered in \cite{Baccelli2018}. For other specific examples when cooperative dynamic systems arise in the analysis of scheduling in communication systems see, e.g., \cite{KW2004,LS2016}.
\end{remark}

We also present a different proof of Theorem \ref{thm:stab_csma_single}, which is specific to our model and is based on a global optimality of the rates $\overline{\varphi}$. We think that this optimality is interesting on its own as it is an important structural property of the rates, and as it provides an example of a decentralised algorithm which maximises a global utility function. We also present a simple proof showing stability of algorithms maximising utility functions over a set, without requiring convexity of the set.

The remainder of this second proof consists of the following steps, which correspond to two lemmas below:
\begin{enumerate} 
\item We show that the FSPs of this model are such that the ``service rates'' the nodes receive are utility maximising (in fact $2$-fair) in the set $\C$. 
This property is proved in Lemma \ref{lemma:fairness_csma};
\item The property \eqn{eq:fluid_stability} of the FSPs (which is sometimes referred to as stability of FSPs), 
and hence the stability of the underlying Markov chain, follows from the utility-maximisation property of the ``service rates.'' We only need to note that this fact is usually proved for convex sets of possible rates, whereas our set $\C$ is not convex. However, convexity is in fact not needed in the stability proof, and we provide a proof for any sets, based on the proof of \cite[Theorem 2]{St2005alq}; this is done in Lemma \ref{lemma:fairness_stability}.
\end{enumerate}

\begin{lemma} \label{lemma:fairness_csma}
For any $\overline x$ with $x_i>0$ for all $i$, the rates $\overline \varphi = \overline \varphi (\overline x)$
 are $2$-fair in the set $\C$ (see the Introduction of this paper or, e.g. \cite{Bonald2001} for definition of $\alpha$-fairness).
\end{lemma}

{\bf Proof of Lemma \ref{lemma:fairness_csma}.} Indeed, due to the definition of the set $\C$, for any $\overline{\mu} \in \C$,
$$
\sum_i x_i \left(\frac{\mu_i}{x_i}\right)^{-1} \ge \sum_i x_i \left(\frac{p_i}{(\sum_{j \in \mathcal{N}_i} p_j) x_i}\right)^{-1}
$$
for the corresponding vector $\overline{p}$. Hence, it is sufficient to show that
$$
\sum_i x_i \left(\frac{\varphi_i}{x_i}\right)^{-1} \le \sum_i x_i \left(\frac{p_i}{(\sum_{j \in \mathcal{N}_i} p_j) x_i}\right)^{-1}
$$
for all vectors $\overline{p}$.

Note that the LHS of the above is equal to $\sum_i x_i \sum_{j \in \mathcal{N}_i} x_j = \sum_i x_i^2 + \sum_i \sum_{j \in \mathcal{N}_i, j \neq i} x_i x_j$. Consider now
\begin{align*}
\sum_i x_i \left(\frac{p_i}{(\sum_{j \in \mathcal{N}_i} p_j) x_i}\right)^{-1} & = \sum_i x_i^2 \left(1+ \sum_{j \in \mathcal{N}_i, j \neq i} \frac{p_j}{p_i}\right) \\ & = \sum_i x_i^2 + \frac{1}{2} \sum_i \sum_{j \in \mathcal{N}_i, j \neq i} \left(x_i^2 \frac{p_j}{p_i} + x_j^2 \frac{p_i}{p_j}\right).
\end{align*}
For any $i$ and $j$,
$$
x_i^2 \frac{p_j}{p_i} + x_j^2 \frac{p_i}{p_j} \ge 2 x_i x_j,
$$
and the equality is possible if and only if $x_i^2 \frac{p_j}{p_i} = x_j^2 \frac{p_i}{p_j}$, which is equivalent to $\frac{p_i}{x_i} = \frac{p_j}{x_j}$. Therefore we obtain
$$
\sum_i x_i \left(\frac{\sum_{j \in \mathcal{N}_i} p_j}{p_i x_i}\right)^{-1} \ge \sum_i x_i^2 + \sum_i \sum_{j \in \mathcal{N}_i, j \neq i} x_i x_j,
$$
and the equality is possible if and only if $\frac{p_i}{x_i} = \frac{p_j}{x_j}$ for all $i$ and $j$. This implies that $\frac{p_i}{x_i}$ has to be a constant for each $i$, as the graph is connected. This concludes the proof of Lemma \ref{lemma:fairness_csma}. \qed

The result of Theorem \ref{thm:stab_csma_single} now follows from stability of FSPs under $\alpha$-fair algorithms (see, \cite{Bonald2001,DeVeciana2001}). One only needs to note that such proofs are usually given for convex sets, but convexity is not in fact needed, and stability may be proved following the lines of the proof of Theorem 2 in \cite[Section 8]{St2005alq}, where it was given in the case $\alpha=1$.
The proof for far more general rate allocations is essentially the same and we provide it here for completeness.

\begin{lemma} \label{lemma:fairness_stability} Let $\C$ be a compact coordinate-convex subset of $\mathbb{R}_+^N$.
Let $h_i: [0,\infty) \to \mathbb{R}$, for each $i$, be an increasing differentiable concave function (the case when $h_i(y)\downarrow -\infty$ as $y\downarrow 0$ is allowed). Let  $g_i:[0,\infty) \to [0,\infty)$, for each $i$, be a continuous non-decreasing function such that $g_i(0)\ge 0$ and $g_i(y) >0$ for $y > 0$.

Consider a family of Lipschitz trajectories $\overline x(t), ~t\ge 0$, in $\mathbb{R}_+^N$,
which satisfy the following conditions:
\beql{eq-gen_sh1}
x_i(t) > 0 \quad \text{implies} \quad  x'_i(t) = \lambda_i - \psi_i(\overline{x}(t)), ~~~\mbox{for almost all $t\ge 0$},
\end{equation}
\beql{eq-gen_sh2}
\sum_i x_i(t) > 0  \quad \text{implies} \quad \mbox{$x_i(\tau) > 0,$ for all $i,$ for all $\tau>t$ sufficiently close to $t$},
\end{equation}
where the rates $\overline{\psi}$ satisfy
\begin{equation} \label{eq:gen}
\overline{\psi} \in \arg\max_{\overline{\mu} \in \C} \sum_i g_i(x_i) h_i(\mu_i).
\end{equation}
Assume that $\overline{\lambda}$ is such that $\overline{\lambda} < \overline{\nu}$ for some $\overline{\nu} \in \C$. 
Then for any constants $0<\delta < K < \infty$, there exists $T>0$ such that, for any such trajectory with $\|x(0)\| = K$,
$$
\|x(T)\| \le \delta.
$$
\end{lemma}

{\bf Proof of Lemma \ref{lemma:fairness_stability}.} In this proof, we will drop the index $t$ for ease of notation. We will also write simply $\psi_i$, with its dependence on $\overline x(t)$ being implicit. 

Property \eqn{eq-gen_sh2} implies that $x_i(t)>0,$ for all $i$, for $t \in (0,\theta)$, where $\theta$ is the first time (if any) when all $x_i(t)$ ``hit'' $0$ simultaneously. Consider a trajectory $\overline x(\cdot)$ in this interval $(0,\theta)$.

Note that there exists $\varepsilon > 0$ such that $\lambda_i < \nu_i - \varepsilon$ for each $i$. Denote by
$$
F(\overline{y}) = \sum_{i=1}^N G_i(y_i) h'_i(\nu_i)
$$
with $G_i(z) = \int_0^z g_i(s) ds$, and note that
\begin{align} \label{eq:Lyapunov_derivative}
(F(\overline{x}))' & = \sum_{i=1}^N h'_i(\nu_i) g_i(x_i) (\lambda_i - \psi_i) \notag \\ & = \sum_{i=1}^N h'_i(\nu_i) g_i(x_i) (\lambda_i - \nu_i) + \sum_{i=1}^N h'_i(\nu_i) g_i(x_i) (\nu_i - \psi_i) \notag \\ & < -\varepsilon \sum_{i=1}^N h'_i(\nu_i) g_i(x_i) + \sum_{i=1}^N h'_i(\nu_i) g_i(x_i) (\nu_i - \psi_i).
\end{align}
As $\overline{\nu} \in C$, due to \eqref{eq:gen},
$$
\sum_{i=1}^N g_i(x_i) h_i(\psi_i) \ge \sum_{i=1}^N g_i(x_i) h_i(\nu_i).
$$
Using this and the concavity of the functions $h_i$, we have
\begin{equation*}
0 \le \sum_{i=1}^N g_i(x_i) \left(h_i(\psi_i) - h_i(\nu_i)\right) \le
\sum_{i=1}^N g_i(x_i) h'_i(\nu_i) (\psi_i - \nu_i).
\end{equation*}
This, together, with \eqref{eq:Lyapunov_derivative}, implies that
$$
(F(\overline{x}))' < - \varepsilon \sum_{i=1}^N h'_i(\nu_i) g_i(x_i).
$$
We can now conclude that $(F(\overline{x}))' $ is negative, bounded away from $0$, as long as $F(\overline{x})$ is positive, bounded away from $0$. Indeed, assume $F(\overline{x}) \ge c_1 > 0$. Then there exists $i$ such that $h_i'(\nu_i) G_i(x_i) \ge c_1/N$ and hence $G_i(x_i) \ge c_2 > 0$ with obvious notation for $c_2$. Therefore, $x_i \ge G_i^{-1}(c_2)$, where $G_i^{-1}$ is the inverse of $G_i$, which is a well defined strictly increasing function due to our conditions on functions $g_i$. Then
$$
(F(\overline{x}))' < - \varepsilon \sum_{i=1}^N h'_i(\nu_i) g_i(x_i) \le - \varepsilon h'_i(\nu_i) g_i(x_i) \le - \varepsilon h'_i(\nu_i) g_i(G_i^{-1}(c_2)) = c_3 < 0,
$$
as $g_i$ is non-decreasing.
This  concludes the proof.
\qed

It is easy to see that the condition \eqref{eq-sh2} implies \eqref{eq-gen_sh2} as the graph $\mathcal{G}$ is connected. Indeed, assume that $x_i(t) = 0$ for some $i$ but $\sum_i x_i(t) > 0$. Let us first assume that there exists $j \in \mathcal{N}_i$ such that $x_j(t) > 0$. Then, due to condition \eqref{eq-sh2}, $x_i(\tau) > 0$ for all $\tau$ sufficiently close to $t$. If $x_j(t) = 0$ for all $j \in \mathcal{N}_i$, due to the fact that $\mathcal{G}$ is connected, there exists $v$ such that $x_v(t) > 0$ and such that there exists a path $i=v_1, v_2,\ldots,v_l=v$ between nodes $i$ and $v$ (i.e. there is an edge between $v_d$ and $v_{d+1}$ for all $d=1,\ldots,l-1$). We can then use the argument above to show that $x_{v_d}(\tau) > 0$ for all $\tau$ sufficiently close to $t$, for all $d$.

The result of Theorem \ref{thm:stab_csma_single} now follows if we take $g_i(y) = y^2$ and $h_i(y) = - y^{-1}$.

\subsection{Proof of Theorem~\ref{thm:rate-stab}}
\label{sec-infinite}

Recall that Theorem~\ref{thm:rate-stab} is for the infinite, as well as finite, system.
It suffices to prove the following.

\begin{lemma}
\label{lem-infinite}
Any FSP starting from zero initial state, $x_i(0)=0$ for all $i$, stays in zero state at all times, $x_i(t)=0, ~t\ge 0$ for all $i$.
\end{lemma}

{\bf Proof of Lemma \ref{lem-infinite}.} Consider any FSP with zero initial state. 
Denote $s(t) = \sup_i x_i(t)/p_i$. This function is Lipschitz, because all $x_i(\cdot)$ are uniformly Lipschitz
and all $0<c \le p_i \le 1$. Time points $t$ where all derivatives $x'_i(t)$ and $s'(t)$ exist, are called regular.
Almost all points $t$ (with respect to Lebesgue measure) are regular.
 We will show that at any regular point $t$, $s'(t)\le 0$.
This will imply that $s(\cdot)$ cannot escape from $0$. Suppose not, and at some regular point $t$, $s'(t)=\eta>0$.  If this is true, 
then there exists a positive function $\delta_1 =\delta_1(\delta) \downarrow 0$ as $\delta \downarrow 0$, such that 
the following holds for any sufficiently small $\delta>0$: (a) there exists $i$ such that the increment 
\beql{eq-incr}
x_i(t+\delta)/p_i-x_i(t)/p_i \ge (\eta/2) \delta,
\end{equation}
(b) $|x_i(\xi)/p_i - s(t)| < \delta_1$ for all $\xi \in [t,t+\delta]$, (c) $x_j(\xi)/p_j - s(t) < \delta_1$ for all $j\in \mathcal{N}_i$ and all
$\xi \in [t,t+\delta]$. If we consider such an $i$, we observe that for any regular $\xi \in [t,t+\delta]$, thanks to properties (b) and (c),
\begin{align*}
\varphi_i(\bar{x}) 
> \frac{(s(t)-\delta_1)p_i}{\sum_{j\in \mathcal{N}_i} (s(t)+\delta_1)p_j} = \frac{s(t)-\delta_1}{s(t)+\delta_1} \varphi_i(\bar{p}).
\end{align*}
Hence, as $\lambda_i < \varphi_i(\bar{p})$, we see that $x'_i(\xi)/p_i \le \epsilon=\epsilon(\delta_1)$, where $\epsilon(\delta_1)\downarrow 0$ as $\delta_1\downarrow 0$. 
Therefore, for a sufficiently small $\delta$ and a corresponding $i$, $x'_i(\xi)/p_i \le \eta/3$ for all regular $\xi \in [t,t+\delta]$.
This contradicts \eqn{eq-incr}.
\qed

\section{Symmetric multi-hop networks with Geometric service requirements: Proof of Theorem~\ref{thm:stability_csma_multi}} \label{sec:multi}

As in the single-hop case, we first present a lemma on the conditions any FSP satisfies. 

\begin{lemma} \label{lemma:fluid_multi}
Any FSP for a multi-hop symmetric network satisfies the following conditions:
\begin{align}
& x_i(t) > 0  \quad \text{implies} \notag \\ & x'_i(t) = \lambda_i - \varphi_i(\overline{x}(t)) + \frac{1-1/k}{m-1} \sum_{j \in \N_i, j \neq i} \varphi_j(\overline{x}(t)), ~ \mbox{for almost all $t\ge 0$}, \label{eq-mh1} \\
& [x_i(t) = 0 ~\mbox{and} ~\sum_{j \in \N_i} x_j(t) > 0]  \quad \text{implies} \quad \frac{d_l^+}{dt} x_i(t) \ge \lambda_i. \label{eq-mh2}
\end{align}

In particular, by property \eqn{eq-mh2}, any FSP for a {\em finite} network is such that $\sum_i x_i(t) > 0$ implies that
$x_i(\tau) > 0, ~\forall i,$ for all $\tau>t$ sufficiently close to $t$.
\end{lemma}

Once again,  a proof of Lemma~\ref{lemma:fluid_multi} may be given following the exact same lines as that of \cite[Theorem 2]{St2005alq}, for a model close to ours, and we omit it here. The last term on the RHS of \eqref{eq-mh1} may be explained as follows: from each $j \in \mathcal{N}_i$, $j \neq i$ messages leave at rate $\varphi_j$, a proportion $1-1/k$ of these do not leave the system, and a further proportion $1/(m-1)$ of those that do not leave the system choose node $i$ as their destination.

We will omit the index $t$ in the remainder of the proof. Fix $\varepsilon > 0$ such that $\lambda + \varepsilon < 1/m$ and consider the Lyapunov function $\frac{1}{2}\sum_i x_i^2$, whose drift is
\begin{align*}
& \sum_i x_i \left(\frac{\lambda}{k} - \varphi_i + \sum_{j \in \mathcal{N}_i, j \neq i} (1-1/k)\frac{1}{m-1} \varphi_j \right) 
\\ & = \frac{\lambda}{k} \sum_i x_i - \frac{1}{k} \sum x_i \varphi_i + \sum_i x_i \left(-(1-1/k) \varphi_i + \sum_{j \in \mathcal{N}_i, j \neq i} (1-1/k)\frac{1}{m-1} \varphi_j\right)
\\ & < - \frac{\varepsilon}{k}\sum_i x_i +  \frac{1}{k} \left(\frac{1}{m} \sum_i x_i - \sum x_i \varphi_i\right) \\ & + (1-1/k) \sum_i \varphi_i \left(-x_i + \frac{1}{m-1} \sum_{j \in \mathcal{N}_i, j \neq i} x_j \right)
\\ & = - \frac{\varepsilon}{k}\sum_i x_i + \frac{1}{k} \left(\frac{1}{m} \sum_i x_i - \sum x_i \varphi_i\right) \\ & + (1-1/k) \sum_i \varphi_i \left(-x_i + \frac{1}{m-1} \left(\frac{x_i}{\varphi_i} - x_i \right) \right)
\\  & = - \frac{\varepsilon}{k}\sum_i x_i + \frac{1}{k} \left(\frac{1}{m} \sum_i x_i - \sum x_i \varphi_i\right) \\ & + (1-1/k) \left( - \frac{m}{m-1} \sum_i x_i \varphi_i + \frac{1}{m-1} \sum_i x_i \right)
\\ & = - \frac{\varepsilon}{k}\sum_i x_i + \left(\frac{1}{k} + (1-1/k) \frac{m}{m-1} \right) \left(\frac{1}{m} \sum_i x_i - \sum x_i \varphi_i\right) 
\\ & = - \frac{\varepsilon}{k}\sum_i x_i + \frac{1}{m-1}\left(m - \frac{1}{k}\right) \left(\frac{1}{m} \sum_i x_i - \sum x_i \varphi_i\right).
\end{align*}
Let $S = \sum_i x_i$. Note that
$$
\frac{1}{S} \sum_i x_i \varphi_i = \sum_i \frac{x_i}{S} \frac{1}{\frac{\sum_{j \in \mathcal{N}_i} x_j}{x_i}} \ge \frac{1}{\sum_i \frac{x_i}{S} \frac{\sum_{j \in \mathcal{N}_i} x_j}{x_i}} = \frac{S}{\sum_i \sum_{j \in \mathcal{N}_i} x_j} = \frac{1}{m},
$$
due to convexity of the function $1/x$. It now follows that
$$
\left(\frac{1}{2} \sum_i x_i^2\right)' < - \frac{\varepsilon}{k}\sum_i x_i.
$$
From here the FSP property \eqn{eq:fluid_stability}, and then Theorem~\ref{thm:stability_csma_multi}, follows.
\qed

%

\section{Open problems} \label{sec:open}

Our main result in the multi-hop setting only concerns symmetric (in terms of arrival intensities as well as routing) networks. We expect similar results to hold in greater generality. An interesting example is the following: assume that the graph is a circle, arrival intensities into each node are constant and equal to $\lambda/k$ but the routing is not symmetric. Upon successful transmission, a message leaves the system with probability $1/k$ or goes to its neighbour on the right with probability $1-1/k$. Rate-balance equations imply that the total workload of each node is $\lambda$ and we expect that the condition $\lambda<1/3$ guarantees stability in this model, as well as in the model with symmetric routing covered by Theorem \ref{thm:stability_csma_multi}. In fact, we conjecture that the same Lyapunov function as the one used in the proof of Theorem \ref{thm:stability_csma_multi} has a negative drift in this scenario as well. Simple calculus shows that to prove this, one needs to show that
$$
\sum_{i=1}^N \frac{x_i (x_i - x_{i+1})}{x_{i-1}+x_i+x_{i+1}} \ge 0
$$
for all vectors $\overline{x}$, with the conventions that $x_{0} = x_N$ and $x_{N+1} = x_1$.

We have ample numerical evidence in support of this hypothesis but currently lack a proof. If the inequality above is proved, it will imply, furthermore, that condition $\lambda < 1/3$ guarantees stability on the circle topology with arrival intensities equal to $\lambda/k$ for each node, every message leaving the system upon a successful transmission with probability $1/k$ and for arbitrary (but same for all nodes) routing to neighbours if the message does not leave the system.

\end{document}